\documentclass[11pt]{article}
\usepackage[utf8]{inputenc}
\usepackage{ amsmath, amsthm, amssymb }
\usepackage{adjustbox}

\usepackage{color}
\usepackage{cite}

\newcommand{\be} { \begin{equation} }
\newcommand{\ee} { \end{equation} }

\newcommand{\comment}[2]{#2}

\allowdisplaybreaks

\newcommand{\eps} {\epsilon}
\begin{document}
\setlength{\unitlength}{1.3cm}
\begin{titlepage}
\vspace*{-1cm}
\begin{flushright}
CERN-TH-2019-087,ZU-TH 30/19
\end{flushright}
\vskip 3.5cm
\begin{center}
\boldmath

{\Large\bf Physical projectors for multi-leg helicity amplitudes\\[3mm] }
\unboldmath
\vskip 1.cm
{\large Tiziano Peraro}$^{a,}$
\footnote{{\tt e-mail: peraro@physik.uzh.ch}} and
{\large Lorenzo Tancredi}$^{b,}$
\footnote{{\tt e-mail: lorenzo.tancredi@cern.ch}}
\vskip .7cm
{\it $^a$ Physik-Institut, Universit\"{a}t Z\"{u}rich, Wintherturerstrasse 190, CH-8057 Z\"{u}rich, Switzerland} \\
{\it $^b$ TH Department, CERN
1 Esplanade des Particules,
Geneva 23,
CH-1211,
Switzerland}
\end{center}
\vskip 2.6cm

\begin{abstract}

  We present a method for building physical projector operators for
  multi-leg helicity amplitudes.  For any helicity configuration of
  the external particles, we define a physical projector which singles
  out the corresponding helicity amplitude.  For processes with more than
  four external legs, these physical projectors
  depend on significantly fewer tensor structures and exhibit a
  remarkable simplicity compared with projector
  operators defined with traditional approaches.
  As an example, we present
  analytic formulas for a complete set of projectors for five-gluon
  scattering.  These have been validated by reproducing known results
  for five-gluon amplitudes up to one-loop.

\end{abstract}
\vfill
\end{titlepage}
\newpage


\section{Introduction}
\label{sec:intro} \setcounter{equation}{0}
\numberwithin{equation}{section}

After the discovery of the Higgs boson, particle physics has entered a new, unprecedented phase, at least in
the recent decades.
In spite of the fact that only a small fraction of the expected full LHC data set has been analysed,
we have already been able to confirm the Standard Model of Particle Physics
as the correct theory of Nature with unprecedented precision  and for energies that span
an impressively large number of orders of magnitude.
Given this state of affairs, the apparent
absence of clear signs of new physics has pushed the particle physics
community into a new era of precision physics.
Indeed, by comparing precise theoretical predictions for suitable observables
with equally precise experimental results, the discovery potential of the LHC can be pushed to energies
beyond its direct reach, increasing our chances to spot elusive signs of new physics phenomena.

Among the ingredients required to produce precise theoretical predictions for complex observables at the LHC,
the calculation of scattering amplitudes for multi-particle final state processes has an important place.
In order to match the experimental precision of many measurements at the LHC,
 two-loop corrections for several processes with up to
five external particles are required.
While we have a quite robust understanding of how such calculations should be performed in
perturbative quantum field theory, their technical complexity constitutes often
a major bottleneck.
Indeed, in spite of the many advancements which have already
brought many previously impossible calculations within reach, our current technology to treat multi-loop and
multi-leg processes has only recently obtained its first results for
processes with more than four external legs and more work is needed to
generalise these to other processes.

There are  many reasons why these processes are difficult.
In this paper, we focus in particular on one of them, which has to do with the way we usually
approach the calculation of scattering amplitudes at loop orders higher than one.
In fact, while at one loop
the underlying simplicity of the scattering amplitudes
has allowed to develop techniques and automated tools to deal with these calculations,
their generalisation
to higher-loop orders has revealed to be quite non-trivial.  The
methods of
integrand reduction~\cite{Ossola:2006us,Giele:2008ve} and generalized unitarity~\cite{Bern:1994zx,Bern:1994cg,Britto:2004nc,Ellis:2007br} have
been extended beyond one loop~\cite{Mastrolia:2011pr,Badger:2012dp,Zhang:2012ce,Mastrolia:2012an,Ita:2015tya} and used to obtain the first analytic
results for two-loop five parton amplitudes in the planar limit~\cite{Badger:2018enw,Abreu:2018zmy,Abreu:2019odu}.  Very recently, the first non-planar five-point two-loop amplitudes have also been published~\cite{Chicherin:2018yne,Chicherin:2019xeg,Badger:2019djh}.  In spite of this, a lot of progress
will be required before these ideas can be applied generally to any class of processes.

An alternative approach to compute multi-loop scattering amplitudes, which is in principle
entirely general and can be applied to any process at any perturbative order,
consists in identifying Lorentz-invariant
form factors, which can, in turn, be extracted from the relevant Feynman diagrams through
suitable projector operators.
This method has proven to be very successful in the calculation of a large number of lower-point (i.e. up to four
external particles) scattering amplitudes up to two and three loops in perturbative quantum field theory.
In spite of being very general, its applicability to multi-leg processes has been hindered
by the increasing complexity of the relevant projection operators when more than four external particles
are considered.
The idea behind this method is very simple. Given the scattering on $n$ particles of different spin,
one parametrises the
corresponding scattering amplitude in terms of a combination of scalar \emph{form factors}
which multiply all possible tensor structures compatible with the symmetries
of the process under consideration. Since the tensors form a basis,
for each of these form factors a projection operator can be defined
as a linear combination of the same tensors, whose coefficients are fixed requiring that the projector
singles out the correct form factor.
Such tensor structures are interpreted as generic $d$-dimensional
objects and all manipulations are performed in conventional dimensional regularisation (CDR).
Clearly, the larger the number of external particles grows, the more independent tensors
have to be included, such that going from four to five external legs typically implies a jump in one
order of magnitude in the number of tensors needed and, therefore, in the corresponding form factors.
Moreover, deriving the projectors requires in general to solve a dense linear system of equations, with as many
equations as the number of independent tensor structures. Solving this system becomes very
soon impractical with conventional computer algebra systems.  Even if
the solution can often  be easily found using alternative approaches
(for example, finite fields and multivariate reconstruction~\cite{vonManteuffel:2015gxa,Peraro:2016wsq,Peraro:2019svx}),
the final result will, in general, be extremely cumbersome, making its practical utility unclear.
For these reasons, while progress has recently been made in defining the required projectors
for the case of five-gluon scattering~\cite{Boels:2018nrr},
their use  for generic five- and higher-point scattering amplitudes
is commonly considered to be a very difficult endeavour.

In this paper, we will show that this is not necessarily the case.
Indeed, while the tensor decomposition described above
implies that external particles are taken to be $d$-dimensional,
one of the things that modern techniques have taught us is that
substantial simplifications happen when helicity amplitudes with only physical four-dimensional external states are considered.
Starting from this insight, in this paper we will show that by fixing the
helicities of the external states, one can
define a set of \emph{physical projectors} which single
out at once the corresponding helicity amplitudes. In general, there will be as many
physical projectors as many independent helicity amplitudes and
each of these projectors will be expressed as a linear combination of the original tensors,
with rational coefficients that depend on the kinematic invariants.
As a matter of fact, for processe with more than four external legs, the number of physical projectors
will typically be much smaller than the original number of $d$-dimensional
ones.  Moreover, in these cases, when expressed  in terms of the original tensors,
only a subset of them will contribute and their number will correspond exactly to the number
of independent helicity amplitudes in the process under consideration.
Finally, the corresponding coefficients will be orders
of magnitude simpler than the ones of the original projectors.
In order words, the approach described in this paper allows us
to get rid at once of all spurious tensor structures
which correspond to the extra $(d-4)$ unphysical degrees of freedom associated to the external states
and to define extremely compact projector operators that single out directly the physical degrees of freedom
from the corresponding scattering amplitudes. To demonstrate the effectiveness of the new projectors,
we will apply them to the calculation of one-loop corrections to five-gluon scattering in QCD.
Recently, an alternative approach, which also exploits the simplifications coming from
four-dimensional external states, has been proposed in~\cite{Chen:2019wyb} and we will
comment more later about differences and similarities to our method.

The rest of the paper is organised as follows. In Section~\ref{sec:standard}, we start by recalling how
the standard $d$-dimensional projectors work and elucidate the shortcomings of the standard approach.
Inspired by this, in Section~\ref{sec:idea}
we illustrate how to define physical projectors which overcome most of these issues.
In Section~\ref{sec:applications}  we use these ideas in order to build a complete set of
physical projectors for the scattering of five gluons in QCD. We
then apply explicitly the newly derived projectors to the
calculation of one-loop corrections to five-gluon scattering in QCD. We stress here that
such calculation is usually deemed to be impractical already at one-loop order with the use of standard
$d$-dimensional projectors. Our approach, instead, allows to complete the analytic
calculations of the one-loop corrections in a few hours on an average
laptop computer, by resorting only to standard computer algebra systems as FORM~\cite{Vermaseren:2000nd}
and Reduze~\cite{vonManteuffel:2012np}.
Finally, we draw our conclusions in Section~\ref{sec:conclusions}.




\section{Shortcomings of the standard approach}
\label{sec:standard} \setcounter{equation}{0}
\numberwithin{equation}{section}
Before discussing the general idea behind the definition of physical projectors, we remind the reader
of the way standard $d$-dimensional projectors work. We will stress why they are so useful in the context
of multiloop calculations and, at the same time, highlight the
shortcomings of the traditional approach.

Typically, multiloop calculations start with the enumeration of the Feynman diagrams
which contribute to the process considered at the corresponding perturbative order.
Diagrams always take the form of a multiple integral
over the momenta of the virtual particles running in the loops, whose integrand is  given
by a rational function in the scalar products among the loop
momenta, the momenta of the external particles and all their polarization structures (polarization
vectors, spin-chains, etc).
By factoring out all loop-independent tensor structures,
one is then left with a large combination of tensor integrals, which are notoriously
very difficult to
compute as long as the tensor indices are not contracted.
 On the contrary,
many effective tools are available
for the calculation of so-called \emph{scalar Feynman integrals}, where all  loop momenta
are contracted either among themselves or with the momenta of the external particles.
Indeed, large numbers of linear relations among these integrals can be derived, among which
the most prominent role is played by the so-called \emph{integration-by-part identities} (IBPs)~\cite{Tkachov:1981wb,Chetyrkin:1981qh,Laporta:2001dd}.
By solving these identities, the large majority of these integrals
can be expressed in terms of a much smaller subset of independent \emph{master integrals}.
Moreover, the very same IBPs allow one to derive differential equations satisfied by the master integrals~\cite{Kotikov:1990kg,Remiddi:1997ny,Gehrmann:1999as},
which are typically much simpler to solve compared to attempting their direct integration
over the loop momenta.
While these steps are in general not straightforward, a lot of progress has been recently made
in their systematisation~\cite{Henn:2013pwa,Primo:2016ebd,Frellesvig:2017aai,Bosma:2017ens,Primo:2017ipr,Mastrolia:2018uzb,Frellesvig:2019kgj}, and we will ignore this aspect in this paper.

Instead, we will focus on the previous step, i.e. on the manipulations required to go from the tensor integrals
stemming from the Feynman diagrams, to the corresponding scalar integrals for which the technology above
can be applied.
Different possible solutions to this problem exist and
in what follows we will focus on one possible
approach. This consists in deriving suitable \emph{projection operators} which,
once applied on the relevant Feynman diagrams, allow one to project out
the required combinations of scalar Feynman integrals in terms of
\emph{scalar form factors} from the overall, non-perturbative,
Lorentz and Dirac tensor structures.
To fix the notation, let us consider the scattering of $n$ spin-$1$ vector bosons, which we assume
to be all outgoing for definiteness, i.e.
\begin{equation}
0 \to V_1(p_1) + ... + V_n(p_n). \label{eq:genproc}
\end{equation}
While working with spin-1 particles
will allow us to reduce the clutter in the notation, it should be clear that the inclusion of
external particles with different spin (scalar, spinors, etc)
would not change any of the conclusions of the following discussion.

We start by observing that Lorentz invariance alone
requires that the scattering amplitude for~\eqref{eq:genproc}
can be schematically written as
\begin{equation}
\mathcal{A}(p_1,...,p_{n-1}) = \eps_1^{\mu_1} ... \eps_n^{\mu_n}\, \mathcal{T}_{\mu_1,...,\mu_n}(p_1,...,p_{n-1})\,,
\label{eq:ampldecomp}
\end{equation}
where $\eps_j^{\mu_j} = \eps_j^{\mu_j}(p_j)$ are the polarization vectors associated to the external bosons
and $\mathcal{T}_{\mu_1,...,\mu_n}(p_1,...,p_n)$ is a rank-$n$ Lorentz
tensor.
This tensor may, in turn, be decomposed into a tensor basis compatible
with the symmetries of the underlying theory and gauge invariance
\begin{equation}
\mathcal{T}^{\mu_1,...,\mu_n}(p_1,...,p_{n-1}) = \sum_{j=1}^M \mathcal{F}_j\, T_j^{\mu_1,...,\mu_n}\,,\label{eq:gendecomp2}
\end{equation}
where the $\mathcal{F}_j$ are scalar form factors.
As it should be easy to realise, the number of independent tensors  $M$
increases extremely fast with the number of external legs. While their exact number depends on whether the external particles
are massless or massive, one can easily go from a handful of tensors for $3$ external bosons, to $\mathcal{O}(10)$ for $4$,
up to $\mathcal{O}(100)$ for $5$ and so on.

Each of the form factors $\mathcal{F}_j$ can then be extracted by applying
a suitably defined projection operator $\mathcal{P}_j$ on the Feynman diagrams
which contribute to the scattering amplitude in the desired theory and to the desired perturbative order.
To derive the projectors we use the fact that the $M$ tensors are a basis, and write
each projector
as a linear combination of the same tensors, contracted with the respective polarisation structures:
\begin{equation}
\mathcal{P}_j = \epsilon^*_{1\, \mu_1}...\epsilon^*_{n\, \mu_n}\,
\sum_{k} c_{jk}\, T_k^{\mu_1,...,\mu_n} \,. \label{eq:projgen}
\end{equation}
The coefficients $c_{jk}=c_{jk}(d;p_1,...,p_{n-1}) $ are, in general, rational functions of the
number of space-time dimensions $d$
and of the scalar products among the external momenta $p_j$.  They can be determined by applying
each of the projectors on the decomposition in eq.~\eqref{eq:ampldecomp} and imposing that
\begin{equation}
\sum_{pol}\,\mathcal{P}_j \,  \mathcal{A}(p_1,...,p_{n-1})  = \mathcal{F}_j\,, \label{eq:systemeqs}
\end{equation}
where the sum runs over the polarisations of the external particles.
More explicitly, the $c_{ij}$ can be computed by inverting the following matrix
\begin{equation} \label{eq:systeminverse}
  c^{-1}_{ij} = \left( \sum_{pol} \epsilon^*_{1\, \mu_1}...\epsilon^*_{n\, \mu_n} \eps_{1\, \nu_1} ... \eps_{n\, \nu_n}\,  \right) T_i^{\mu_1,...,\mu_n}  T_j^{\nu_1,...,\nu_n}.
\end{equation}
Notice that, in all the equations so far, the polarization vectors are
treated symbolically.  After summing them over the external
polarization, each $\epsilon^*_{i\, \mu_i}\epsilon_{i\, \nu_i}$ is
replaced by the expression consistent with  the
gauge constraints that have been applied in defining the basis in eq.~\eqref{eq:gendecomp2}.  In this way, the
matrix elements defined in the previous equations are rational
functions of the Mandelstam invariants and the space-time dimensions $d$.

If the matrix in eq.~\eqref{eq:systeminverse} can be inverted,
all form factors can in principle be computed in terms of
scalar Feynman integrals, for which the technology of IBPs and differential equations can be employed.
As a next step, one usually starts from the amplitude in eq.~\eqref{eq:ampldecomp} and
fixes the polarisations of the external states, forcing them
in $d=4$ space-time dimensions. This allows one to define \emph{helicity amplitudes},
which can be written as linear combinations of the $M$ scalar form factors $\mathcal{F}_j$.
We note that this corresponds to working in the 't Hooft-Veltman scheme (tHV), where
external states are taken in $4$ space-time dimensions, while virtual ones are
taken in $d$ continuous dimensions~\cite{tHooft:1972tcz}.

While this construction is clearly very general, it should be equally clear that
finding the solution of  eq.~\eqref{eq:systemeqs} (that allows one
to define the projectors in the first place) can
become highly non-trivial when a large number of tensor structures
is involved. Moreover, even if a solution can be found, the projectors themselves can
become very soon extremely cumbersome, making their practical use
quite difficult. Finally, one might wonder if taking well engineered linear
combinations of tensors (and therefore linear combinations of the
original form factors) as a new basis of
objects could simplify the system in eq.~\eqref{eq:systemeqs} and with it, its final solutions.
Unfortunately, since virtually any linear combination could work equally well, there is no obvious criterion
to select a basis of tensors over any other.

To show how the complexity of this approach can easily get out of hand,
let us consider the prototypical example of the scattering of  five massless spin-1
particles in a parity-invariant theory. Among the others, this covers the case of
five-gluon scattering in QCD. By generating all possible Lorentz structures,
one is left with $1724$ tensors. Imposing that the external gluons are transversely polarised,
i.e. $\eps_j \cdot p_j = 0$ for $j=1,...,5$, and imposing invariance under gauge transformations
(or equivalently fixing the gauge of the external gluons) reduces their number to $142$.
We proceed then by writing the scattering amplitude as
\begin{equation}
\mathcal{A}(p_1,...,p_4) = \epsilon_1^{\mu_1}...\epsilon_5^{\mu_5} \, \sum_{j=1}^{142} \mathcal{F}_j\,
T_j^{\mu_1,...,\mu_5}(p_1,...,p_4)\,. \label{eq:ampl5ggen}
\end{equation}
Following the discussion around eqs.(\ref{eq:projgen},\ref{eq:systemeqs}), we can attempt to derive
the corresponding $142$ projectors
\begin{equation}
\mathcal{P}_j = \epsilon_{1\mu_1}^*...\epsilon_{5\mu_5}^* \,
\sum_{k=1}^{142} c_{jk}(d;p_1,...,p_4)\, T_k^{\mu_1,...,\mu_5}\,. \label{eq:proj5g}
\end{equation}
The corresponding system of equations for the coefficients $c_{jk}$ is
too complicated to be solved by a naive use of Mathematica or
FORM and alternative methods must be considered.  A possible
strategy towards a solution has been outlined in~\cite{Boels:2018nrr}. Another
possibility consists in using techniques based on algebraic
manipulations over finite fields: the system can be solved
numerically modulo prime numbers and the exact
symbolic solution can then be reconstructed from multiple numerical
evaluations (see e.g. refs.~\cite{vonManteuffel:2015gxa,Peraro:2016wsq,Peraro:2019svx}).  While these techniques allow us to get to a
solution quite easily, it is enough to look at the resulting
projectors to understand the limits of this method.  Indeed, the
$142 \times 142$ coefficients $c_{jk}$ in eq.~\eqref{eq:proj5g} occupy
alone $1$Gb of disk space.\footnote{The inversion has been performed
  using \textsc{FiniteFlow}~\cite{Peraro:2019svx}.  The dimension of 1Gb refers to
  the GCD-simplified (but not factorized) analytic result written to a file.
  We stress that this calculation was only done as a test and it is
  not required when using the physical projectors we present in this
  paper. By comparison, the ancillary file attached to this paper 
  contains a full set of physical projectors for the same process in about $750$Kb.}  Having in mind the complexity of the Feynman diagrams
required, for example, to compute the scattering of five gluons
at two loops in QCD, it appears evident that such an approach is
deemed to fail. Moreover, it should as well be clear that the
perspective of using the same approach for even larger numbers of
external particles (for example in the six-gluon case) appears
entirely unfeasible.

Motivated by these problems, in the next section we describe how
most of these limitations can be lifted by defining suitable \emph{physical projector operators} which
single out directly the helicity amplitudes required for the calculation we are interested in.
As we will see, this approach applied to the case of five-gluon scattering will solve at once many problems.
First,
the majority of the $142$ tensor structures in eq.~\eqref{eq:ampl5ggen} will turn out to be  redundant.
Moreover, in comparison with the $1$Gb of data required to specify the
standard projectors, our new physical projector operators will end up being extremely compact
and easy to use.


\section{Physical projectors for helicity amplitudes}
\label{sec:idea} \setcounter{equation}{0}
\numberwithin{equation}{section} In this section, we present the main
result of this paper.  We show that, by projecting directly onto the
helicity amplitudes defined in tHV scheme,\footnote{Our approach
  actually applies to any dimensional regularization scheme where the
  external states are treated in four dimensions.  In particular, the
  projectors built with our method are also valid in the
  Four-Dimensional-Helicity scheme~\cite{Bern:2002zk}, since the
  latter only differs from tHV because of a different treatment of the
  internal gluon states.  In the remainder of this paper, we still
  only refer to the tHV scheme for simplicity.} we can build a set of
\emph{physical projectors} having compact analytic expressions and
involving a substantially smaller number of tensor structures.

Let us consider once more the general decomposition for the scattering amplitude
of $n$ spin-1 bosons in eqs.(\ref{eq:ampldecomp},\ref{eq:gendecomp2}). Once more,
particles of different spin can be accommodated by a straightforward generalisation
of this discussion. Having an explicit representation for the general amplitude, we can imagine
to consider four-dimensional external states and fix their helicities
in all possible ways. We define the total number of helicity amplitudes to be $h_\lambda$.
Clearly, in a case where not all external particles are different,
many of the helicity configurations will not be independent  and may be
 obtained from the independent ones by permutations of the external legs and
complex conjugation. We ignore this detail for now.
If the helicity of the boson $j$ is $\lambda_j$, we write the scattering amplitude
as $\mathcal{A}_{\lambda_1,...,\lambda_n}(p_1,...,p_{n}) $.
In the case of $n$ massless external spin-1 bosons, each particle can have two helicity states, such that
there will be in total $h_\lambda = 2^n$ different helicity amplitudes.
We stress that, while the helicity amplitudes are enough to reconstruct the full
structure of the scattering amplitude, typically their number grows with the number of external particles
much slower than the number
of independent tensors. Indeed, for $5$ massless external spin-1 bosons, there are only $h_\lambda = 32$
independent helicity configurations, in comparison with the $M=142$ different tensor structures discussed
in the previous section. For $6$ external gluons, there are only $h_\lambda = 64$.
Armed with this observation, we would like to define projectors operators which,
instead of projecting on all ``unphysical'' form factors $\mathcal{F}_j$,
project only onto the $h_\lambda$ independent helicity amplitudes.

We first
recall that, for all helicities $\lambda_j$, we can define explicit
polarization states using the spinor-helicity formalism~\cite{Mangano:1987xk,Berends:1987me,Dixon:1996wi,Arkani-Hamed:2017jhn}, in
terms of massless spinors $|j\rangle = |p_j\rangle$ and $|j]=|p_j]$
with negative and positive helicity.  As an
example, polarization vectors $\epsilon_\lambda^\mu$ for massless
bosons can be defined as
\begin{equation} \label{eq:sp4polvec}
  \epsilon_+^\mu(p) = \frac{\langle \eta | \sigma^\mu | p]}{\sqrt{2}\, \langle \eta\, p \rangle}, \qquad \epsilon_-^\mu(p) = \frac{\langle p | \sigma^\mu | \eta]}{\sqrt{2}\, [ p\, \eta ]},
\end{equation}
where $\eta$ is an arbitrary reference vector.  Analogous formulas
exist for polarization states of particles with different spin and
massive particles as well.  Moreover, when one deals with
spinor products, it is often convenient
to work with objects which are invariant under little group scaling
\begin{equation}\label{eq:littlegroup}
  | j \rangle \to t_j\, | j \rangle\,, \quad | j ] \to t_j^{-1} | j]\,.
\end{equation}
It is always possible to
define a rescaled
amplitude $\overline{\mathcal{A}}_{\lambda_1,...,\lambda_n}(p_1,...,p_{n-1})$
which is invariant under little-group scaling
by dividing it by a suitable prefactor $K_{\lambda_1,...,\lambda_n}$
in the spinor products
\begin{equation}
  \overline{\mathcal{A}}_{\lambda_1,...,\lambda_n} = \frac{1}{K_{\lambda_1,...,\lambda_n}}\, \mathcal{A}_{\lambda_1,...,\lambda_n}.
\end{equation}
While there is no unique choice for the prefactor
$K_{\lambda_1,...,\lambda_n}$,  it can be easily built based
on the external helicities and it is independent of the loop order.
Explicit examples will be given in the next section.
The rescaled amplitudes $\overline{\mathcal{A}}_{\lambda_1,...,\lambda_n}$ are by construction
invariant under the little group transformation~\eqref{eq:littlegroup}
and thus independent of any spinor phase.

With these concepts in mind, we start from eqs.~(\ref{eq:ampldecomp},\ref{eq:gendecomp2}) and fix the helicities
of the external particles as
\begin{equation}
\overline{\mathcal{A}}_{\lambda_1,...,\lambda_n}(p_1,...,p_{n-1}) = \frac{1}{K_{\lambda_1,...,\lambda_n}}\, \epsilon_{\lambda_1}^{\mu_1}...
\epsilon_{\lambda_n}^{\mu_n} \sum_{j=1}^M \mathcal{F}_j\, T_j^{\mu_1,...,\mu_n}. \label{eq:helfixed1}
\end{equation}
We stress that, while in the previous sections we treated the external
states mostly symbolically (i.e.\ only in order to yield a sum over
polarizations), here the external polarization states
$\epsilon_{\lambda_j}^{\mu_j}$ are explicit polarizations built out of
spinors.  We can rewrite the previous equation as
\begin{equation} \label{eq:helfixed3}
  \overline{\mathcal{A}}_{\lambda_1,...,\lambda_n}(p_1,...,p_{n-1})
= \sum_{j=1}^M \mathcal{F}_j\,R^{\lambda_1,...,\lambda_n}_j
\end{equation}
where we have defined
\begin{equation}\label{eq:Rj}
  R^{\lambda_1,...,\lambda_n}_j = \frac{1}{K_{\lambda_1,...,\lambda_n}}\,\epsilon_{\lambda_1}^{\mu_1}...
\epsilon_{\lambda_n}^{\mu_n}  \; T_j^{\mu_1,...,\mu_n}.
\end{equation}
Because the objects $R_j$ defined in Eq.~\eqref{eq:Rj} are also
invariant under little group transformations, they can be parametrised
in terms of $3n-10$ invariants $x_j$,
\begin{equation}
  \label{eq:6}
  R^{\lambda_1,...,\lambda_n}_j = R^{\lambda_1,...,\lambda_n}_j(x_1,...,x_{3n-10}).
\end{equation}
These invariants, in turn, can always be chosen such that all scalar quantities
involving spinors and polarization vectors are rational
functions of the $x_j$.  We point out that, for kinematics with
external massive particles, the functions
$R^{\lambda_1,...,\lambda_n}_j$ will also depend on the external
masses.

This allows us to formulate our central result.
In fact, at this point, we simply promote eq.~\eqref{eq:helfixed3} to become a new
\emph{helicity projection
operator} by the formal substitution $\mathcal{F}_j \to \mathcal{P}_j$,
where $\mathcal{P}_j$ is the projector that singles out $\mathcal{F}_j$
as defined in eq.(\ref{eq:projgen},\ref{eq:systemeqs}).
We have defined in this way a set of as many helicity projectors as the number of
independent helicity amplitudes
\begin{equation}
\mathcal{P}_{\lambda_1,...,\lambda_n}
= \sum_{j=1}^M R^{\lambda_1,...,\lambda_n}_j\,\mathcal{P}_j\,. \label{eq:helproj}
\end{equation}
By using eq.~\eqref{eq:projgen} and remembering that all scalar products $p_i\cdot p_j$
can be written as rational functions in the variables $x_j$, we
immediately see that the new
helicity projectors will be also written as a linear combination of the original tensors
\begin{equation}
\mathcal{P}_{\lambda_1,...,\lambda_n}(p_1,...,p_{n-1})
= \epsilon_{1\mu_1}^*...\epsilon_{n\mu_n}^*\, \sum_{k=1}^M \mathcal{C}^{\lambda_1,\ldots,\lambda_n}_{k} T_k^{\mu_1,...,\mu_n}
\label{eq:helproj2}
\end{equation}
where the $\mathcal{C}_{k}=\mathcal{C}^{\lambda_1,\ldots,\lambda_n}_{k}(d;x_1,...,x_{3n-10})$
will be rational functions in $d$ and in the $x_j$.  One should realise here that, while the projectors in eq.~\eqref{eq:helproj2} are defined to
project out the helicity amplitudes $\mathcal{A}_{\lambda_1,...,\lambda_n}$, the
polarisation vectors $\epsilon_j$
that appear on the right-hand side of the previous equation do not
have their polarisation fixed.
On the contrary, as already explained above, they are applied
on the Feynman diagrams by summing over their helicities as described
in eq.~\eqref{eq:systemeqs}.
By construction, these projectors single out the (rescaled) helicity amplitudes
\begin{equation}
\overline{\mathcal{A}}_{\lambda_1,...,\lambda_n} = \sum_{pol}\,\mathcal{P}_{\lambda_1,...,\lambda_n} \,  \mathcal{A}(p_1,...,p_{n-1})\, . \label{eq:physprojamp}
\end{equation}

As hinted to above, in general the sum in eq.~\eqref{eq:helproj2}
should run over all $M$ independent tensors and
 the coefficients $\mathcal{C}_{k}^{\lambda_1,...,\lambda_n}$
 will depend explicitly on the number of space-time dimensions $d$.
Nevertheless, it turns out that for processes with more than four external legs, this is not the case.
In fact, we observe that in order to
compute helicity amplitudes we are allowed to only consider the
projection  of the tensor
$\mathcal{T}_{\mu_1,...,\mu_n}$ defined in eq.~\eqref{eq:ampldecomp}
onto the four-dimensional physical space, where the external
polarisations live.  For processes with five or more external legs,
 this four-dimensional space is spanned by four independent
external momenta.  Hence, we may restrict the tensor basis to span the
physical space defined by four independent external legs only.  This can be
effectively achieved in the decomposition of
eq.~\eqref{eq:gendecomp2}, simply by removing all tensors containing
the metric tensor $g^{\mu \nu}$.  This can also be seen by observing
that in this case we can decompose $g^{\mu \nu}$ into a
four-dimensional part and a $(-2\epsilon)$-dimensional part, as
\begin{equation}
  g^{\mu \nu} = g_{[4]}^{\mu \nu} + g_{[-2\epsilon]}^{\mu \nu} = g_{[-2\epsilon]}^{\mu \nu} + \mathcal{O}(p_i^\mu p_j^\nu), \label{eq:gmunudec}
\end{equation}
where the last equality states that the four dimensional metric tensor
$g_{[4]}^{\mu \nu}$ is a linear combination of tensors
$p_i^\mu p_j^\nu$ built out of the four independent external momenta.  Hence, we are allowed
to replace $g^{\mu\nu}\to g_{[-2\epsilon]}^{\mu \nu}$ in our general tensor
decomposition.  We then observe that tensors with
$g_{[-2\epsilon]}^{\mu \nu}$ are trivially orthogonal to the other
tensors and  the inversion of the matrix in
eq.~\eqref{eq:systeminverse} can be performed separately in the
four-dimensional and in the $(-2 \epsilon)$-dimensional space.
Moreover, all the coefficients
$R^{\lambda_1,...,\lambda_n}_j$ multiplying tensors which depend on
$g_{[-2\epsilon]}^{\mu \nu}$ also vanish by orthogonality.
Putting everything
together, this is effectively equivalent to removing the metric tensor
$g^{\mu\nu}$ from the very beginning in the tensor decomposition.  A
corollary of this observation is the fact that the physical projectors
for these processes are independent of the space-time dimension $d$
(because such a dependence may only come from the metric tensor).  We
stress, again, that this is true only for processes with more than
four external legs.

Let us see what this implies for a generic $n$-point amplitude with $n\geq 5$
vector bosons.  As we just stated, we may build a physical tensor
basis directly from tensor of the form
$  p_{j_1}^{\mu_1}\cdots p_{j_n}^{\mu_n} $,
where all the $p_{j_k}$ are drawn from a subset of four (independent)
external momenta.  In total, we have $4^n$ such combinations.  For spin-1 bosons we can
always impose transversality for each external particle, i.e. $\epsilon_j\cdot p_j=0$, going down
in this way to $3^n$ tensors. Moreover, if the bosons are massless, by
 fixing their gauge, e.g.\ with
the cyclic choice
$\eps_j\cdot p_{j+1}=0$, with $p_{n+1} = p_1$, we are left with a total of $2^n$ independent tensors.
This is consistent with the fact that, for the scattering of $n$ massless spin-1 bosons,
there are $h_\lambda=2^n$ independent helicity amplitudes and we
expect only $2^n$ tensors to be relevant for their reconstruction.
Hence, the inversion needed for computing the physical projectors can
be performed in a (significantly smaller) $h_\lambda$-dimensional
tensor subspace.

We verified this by computing physical projectors in several five-point examples.
In the case of five-gluon scattering (which will be discussed more in detail in the next section), as we saw in Section~\ref{sec:standard}, the general $d$-dimensional tensor structure requires
$142$ tensors after gauge-invariance and transversality conditions have been imposed on the
external gluons, while only the $h_\lambda = 2^5 = 32$ structures of the form
$p_{j_1}^{\mu_1}p_{j_2}^{\mu_2}p_{j_3}^{\mu_3}p_{j_4}^{\mu_4}p_{j_5}^{\mu_5}$
turn out to contribute to the helicity projectors.
Similarly, we also studied the scattering of four gluons and a scalar,
which is relevant for Higgs boson plus two jets production at hadron colliders,
$gg\to H gg$. In this case one has $h_\lambda = 2^4 = 16$ independent helicity amplitudes.
On the other hand, the full $d$-dimensional tensor structure would require $43$
different tensors after gauge-invariance and transversality conditions have been imposed on the
external gluons.  We verified explicitly that in order to project directly over
helicity amplitudes we need, as expected, only
the $16$ tensor structures built out of the $4$ independent external momenta
$p_{j_1}^{\mu_1}p_{j_2}^{\mu_2}p_{j_3}^{\mu_3}p_{j_4}^{\mu_4}$.

To summarise, given these considerations, in order to reconstruct the
helicity projectors defined in equation eq.~\eqref{eq:helproj} one
never needs to go through the whole $d$-dimensional tensor
structure. In practice, if the number of external legs is $n\geq 5$,
we simply reinterpret all formulas above, i.e.
eqs.(\ref{eq:helfixed1},\ref{eq:helfixed3},\ref{eq:helproj},\ref{eq:helproj2})
with $M=h_\lambda$, as the number of the independent tensors in $d=4$
dimensions built from the combinations
$ p_{j_1}^{\mu_1}\cdots p_{j_n}^{\mu_n} $.  This allows us to simplify
even further the derivation of the helicity projectors since, in
eq.~\eqref{eq:systeminverse}, only a $h_\lambda \times h_\lambda$
matrix has to be considered instead of a typically much larger
$M \times M$ one.  One may also perform the inversion
in eq.~\eqref{eq:systeminverse}, either in the full tensor space or in
the physical subspace, numerically over finite fields and reconstruct
the analytic physical projectors directly.
With the latter approach,
using \textsc{FiniteFlow}~\cite{Peraro:2019svx}, the analytic
reconstruction of the physical projectors becomes extremely
efficient.  For five-point processes, it typically takes a few seconds on a modern laptop.

We conclude this section with an observation about the choice of
variables $x_j$, which we will then make more explicit in the next
section with an example.  As already stated, one can choose invariants
which offer a rational parametrization of the spinor components, up to
a little group and a Poincar\'{e} transformation.  With this choice, all
the functions $R^{\lambda_1,...,\lambda_n}_j$ and
$\mathcal{C}_{k}^{\lambda_1,...,\lambda_n}$ above are rational.  A
notable example are momentum twistor variables~\cite{Hodges:2009hk,Badger:2013gxa,Badger:2016uuq}.
Alternatively, one may choose to use Mandelstam invariants instead.
In this case, one can still obtain a unique representation for
$R^{\lambda_1,...,\lambda_n}_j$ and
$\mathcal{C}_{k}^{\lambda_1,...,\lambda_n}$ by identifying a set of
independent square roots and requiring the result to be multilinear in
these square roots.  The coefficients of each independent monomial in
these square roots, which are rational functions of the Mandelstam
invariants, can be treated independently of each other.  As an example,
with massless 5-point kinematics, we have only one square root which
can be identified with the parity odd invariant
$\textrm{tr}_5=\textrm{tr}(\gamma_5\, p_1\,p_2\,p_3\, p_4)$.  Every
little-group invariant function $R$ of the spinor variables can thus
be written in a unique way as $R=R_++\textrm{tr}_5\, R_-$ where the
parity-even and odd components $R_+$ and $R_-$ are rational functions
of Mandelstam invariants.  In practice, in order to obtain such a
representation for our physical projectors, it is often convenient to
first get a rational representation of the
$R^{\lambda_1,...,\lambda_n}_j$ in terms of momentum twistors
variables and then convert it back to Mandelstam invariants, since
this sidesteps the need of performing tedious spinor algebra.  After
that, the parity  even and odd components of
$\mathcal{C}_{k}^{\lambda_1,...,\lambda_n}$ can be computed from the
corresponding ones of $R^{\lambda_1,...,\lambda_n}_j$,
independently of each other.  The same approach easily generalises to the
presence of several independent square roots.

Before going on with an explicit example, it is interesting to compare our method to the one recently proposed in~\cite{Chen:2019wyb},
which also exploits explicitly simplifications coming from taking external particles in
$d=4$ space-time dimensions.
While this is conceptually similar to our approach and we expect the conclusions to be
equivalent, there are some important differences.
In comparison to~\cite{Chen:2019wyb}, we do not need to perform an explicit decomposition
of the external polarisation states in terms of four-dimensional external momenta in order
to see the relevant simplifications in the tensor structure.
Using our approach, our helicity projectors are \emph{uniquely}
written as linear combinations
of standard, $d$-dimensional projector operators. This
allow us to perform all manipulations in the standard tHV scheme, without having to
make sure that our $d$-dimensional regularisation scheme is consistent.
Finally, this different point of view allows us to see straight-away how the
potential of the method can be fully exploited only starting from $n \geq 5$
external particles, where the simplifications to the tensor structure become
more substantial.  As described
above, for these processes, we can immediately
exclude all tensor structures which are not independent, by using the formal
decomposition in eq.~\eqref{eq:gmunudec}.


\section{Physical projectors for five-gluon scattering}
\label{sec:applications} \setcounter{equation}{0}
\numberwithin{equation}{section}

In order to show the potential of the method that we propose, in this section we apply
it to the case of five-gluon scattering in QCD.
In Section~\ref{sec:standard}, we have already pointed to the difficulties in applying standard
$d$-dimensional projectors to the scattering of five gluons.  In particular,
we have shown that a generic tensor decomposition requires 142 independent structures, see eq.~\eqref{eq:ampl5ggen},
and that the corresponding projector operators given in eq.~\eqref{eq:proj5g} appear to be extremely cumbersome.

Let us then start off by considering the scattering of five massless gluons
$$0 \to g(p_1) + g(p_2) + g(p_3) + g(p_4) + g(p_5)\,,$$
with $p_5 = -p_1-p_2-p_3-p_4$ and $p_j^2 = 0$ for $j=1,..,5$.
The amplitude depends on five independent kinematical invariants, which we pick to be $s_{12}$, $s_{23}$, $s_{34}$, $s_{45}$ and $s_{51}$,
where $s_{ij} = (p_i+p_j)^2$. The parity-odd invariant
$$\textrm{tr}_5=\textrm{tr}(\gamma_5\, p_1\,p_2\,p_3\, p_4)$$
will also play an important role in the following discussion.
As already outlined in eq.~\eqref{eq:ampl5ggen}, the most general tensor decomposition of the
scattering amplitude reads
\begin{equation}
\mathcal{A}(p_1,...,p_4) = \epsilon_1^{\mu_1}...\epsilon_5^{\mu_5} \, \sum_{j=1}^{142} \mathcal{F}_j\,
T_j^{\mu_1,...,\mu_5}(p_1,...,p_4)\,, \label{eq:ampl5g}
\end{equation}
where the tensors $T_j^{\mu_1,...,\mu_5}(p_1,...,p_4)$ are built out of the four
independent momenta $p_j^\mu$, $j=1,...,4$ and the metric tensor $g^{\mu\nu}$.
In order to be left with only $142$ tensor structures we have used the fact that $\epsilon_j \cdot p_j = 0$ for $j=1,...,5$
and we have also imposed a cyclic gauge choice on the external gluons as follows
\begin{equation}
\epsilon_1 \cdot p_2 = \epsilon_2 \cdot p_3 = \epsilon_3 \cdot p_4 = \epsilon_4 \cdot p_5 = \epsilon_5 \cdot p_1 = 0\,. \label{eq:gauge5g}
\end{equation}
While fixing the gauge explicitly is not necessary, it is useful to obtain tensors that are as compact as possible.

As argued in detail in the previous section, since the scattering
amplitude depends on $4$ independent momenta and we are interested in
projecting directly on the physical helicity amplitudes, we can drop
all tensors in~\eqref{eq:ampl5g} which depend explicitly on the metric
tensor $g^{\mu\nu}$. In this way we are left, as expected, with the
$32$ tensors
\begin{alignat}{4} \label{eq:tensors5g}
T_1^{\mu_1,...,\mu_5} = {}& p_1^{\mu_2}\, p_1^{\mu_3}\, p_1^{\mu_4}\, p_2^{\mu_5}\, p_3^{\mu_1}\,,\qquad &
T_2^{\mu_1,...,\mu_5} = {}& p_1^{\mu_2}\, p_1^{\mu_3}\, p_2^{\mu_4}\, p_2^{\mu_5}\, p_3^{\mu_1}\,, \nonumber \\
T_3^{\mu_1,...,\mu_5} = {}& p_1^{\mu_2}\, p_1^{\mu_3}\, p_1^{\mu_4}\, p_3^{\mu_1}\, p_3^{\mu_5}\,, \qquad &
T_4^{\mu_1,...,\mu_5} = {}& p_1^{\mu_2}\, p_1^{\mu_3}\, p_2^{\mu_4}\, p_3^{\mu_1}\, p_3^{\mu_5}\,, \nonumber \\
T_5^{\mu_1,...,\mu_5} = {}& p_1^{\mu_2}\, p_1^{\mu_4}\, p_2^{\mu_3}\, p_2^{\mu_5}\, p_3^{\mu_1}\,, \qquad &
T_6^{\mu_1,...,\mu_5} = {}& p_1^{\mu_2}\, p_2^{\mu_3}\, p_2^{\mu_4}\, p_2^{\mu_5}\, p_3^{\mu_1}\,, \nonumber \\
T_7^{\mu_1,...,\mu_5} = {}& p_1^{\mu_2}\, p_1^{\mu_4}\, p_2^{\mu_3}\, p_3^{\mu_1}\, p_3^{\mu_5}\,, \qquad &
T_8^{\mu_1,...,\mu_5} = {}& p_1^{\mu_2}\, p_2^{\mu_3}\, p_2^{\mu_4}\, p_3^{\mu_1}\, p_3^{\mu_5}\,, \nonumber \\
T_9^{\mu_1,...,\mu_5} = {}& p_1^{\mu_3}\, p_1^{\mu_4}\, p_2^{\mu_5}\, p_3^{\mu_1}\, p_4^{\mu_2}\,, \qquad &
T_{10}^{\mu_1,...,\mu_5} = {}& p_1^{\mu_3}\, p_2^{\mu_4}\, p_2^{\mu_5}\, p_3^{\mu_1}\, p_4^{\mu_2}\,, \nonumber \\
T_{11}^{\mu_1,...,\mu_5} = {}& p_1^{\mu_3}\, p_1^{\mu_4}\, p_3^{\mu_1}\, p_3^{\mu_5}\, p_4^{\mu_2}\,, \qquad &
T_{12}^{\mu_1,...,\mu_5} = {}& p_1^{\mu_3}\, p_2^{\mu_4}\, p_3^{\mu_1}\, p_3^{\mu_5}\, p_4^{\mu_2}\,, \nonumber \\
T_{13}^{\mu_1,...,\mu_5} = {}& p_1^{\mu_4}\, p_2^{\mu_3}\, p_2^{\mu_5}\, p_3^{\mu_1}\, p_4^{\mu_2}\,, \qquad &
T_{14}^{\mu_1,...,\mu_5} = {}& p_2^{\mu_3}\, p_2^{\mu_4}\, p_2^{\mu_5}\, p_3^{\mu_1}\, p_4^{\mu_2}\,, \nonumber \\
T_{15}^{\mu_1,...,\mu_5} = {}& p_1^{\mu_4}\, p_2^{\mu_3}\, p_3^{\mu_1}\, p_3^{\mu_5}\, p_4^{\mu_2}\,, \qquad &
T_{16}^{\mu_1,...,\mu_5} = {}& p_2^{\mu_3}\, p_2^{\mu_4}\, p_3^{\mu_1}\, p_3^{\mu_5}\, p_4^{\mu_2}\,, \nonumber \\
T_{17}^{\mu_1,...,\mu_5} = {}& p_1^{\mu_2}\, p_1^{\mu_3}\, p_1^{\mu_4}\, p_2^{\mu_5}\, p_4^{\mu_1}\,, \qquad &
T_{18}^{\mu_1,...,\mu_5} = {}& p_1^{\mu_2}\, p_1^{\mu_3}\, p_2^{\mu_4}\, p_2^{\mu_5}\, p_4^{\mu_1}\,, \nonumber \\
T_{19}^{\mu_1,...,\mu_5} = {}& p_1^{\mu_2}\, p_1^{\mu_3}\, p_1^{\mu_4}\, p_3^{\mu_5}\, p_4^{\mu_1}\,, \qquad &
T_{20}^{\mu_1,...,\mu_5} = {}& p_1^{\mu_2}\, p_1^{\mu_3}\, p_2^{\mu_4}\, p_3^{\mu_5}\, p_4^{\mu_1}\,, \nonumber \\
T_{21}^{\mu_1,...,\mu_5} = {}& p_1^{\mu_2}\, p_1^{\mu_4}\, p_2^{\mu_3}\, p_2^{\mu_5}\, p_4^{\mu_1}\,, \qquad &
T_{22}^{\mu_1,...,\mu_5} = {}& p_1^{\mu_2}\, p_2^{\mu_3}\, p_2^{\mu_4}\, p_2^{\mu_5}\, p_4^{\mu_1}\,, \nonumber \\
T_{23}^{\mu_1,...,\mu_5} = {}& p_1^{\mu_2}\, p_1^{\mu_4}\, p_2^{\mu_3}\, p_3^{\mu_5}\, p_4^{\mu_1}\,, \qquad &
T_{24}^{\mu_1,...,\mu_5} = {}& p_1^{\mu_2}\, p_2^{\mu_3}\, p_2^{\mu_4}\, p_3^{\mu_5}\, p_4^{\mu_1}\,, \nonumber \\
T_{25}^{\mu_1,...,\mu_5} = {}& p_1^{\mu_3}\, p_1^{\mu_4}\, p_2^{\mu_5}\, p_4^{\mu_1}\, p_4^{\mu_2}\,, \qquad &
T_{26}^{\mu_1,...,\mu_5} = {}& p_1^{\mu_3}\, p_2^{\mu_4}\, p_2^{\mu_5}\, p_4^{\mu_1}\, p_4^{\mu_2}\,, \nonumber \\
T_{27}^{\mu_1,...,\mu_5} = {}& p_1^{\mu_3}\, p_1^{\mu_4}\, p_3^{\mu_5}\, p_4^{\mu_1}\, p_4^{\mu_2}\,, \qquad &
T_{28}^{\mu_1,...,\mu_5} = {}& p_1^{\mu_3}\, p_2^{\mu_4}\, p_3^{\mu_5}\, p_4^{\mu_1}\, p_4^{\mu_2}\,, \nonumber \\
T_{29}^{\mu_1,...,\mu_5} = {}& p_1^{\mu_4}\, p_2^{\mu_3}\, p_2^{\mu_5}\, p_4^{\mu_1}\, p_4^{\mu_2}\,, \qquad &
T_{30}^{\mu_1,...,\mu_5} = {}& p_2^{\mu_3}\, p_2^{\mu_4}\, p_2^{\mu_5}\, p_4^{\mu_1}\, p_4^{\mu_2}\,, \nonumber \\
T_{31}^{\mu_1,...,\mu_5} = {}& p_1^{\mu_4}\, p_2^{\mu_3}\, p_3^{\mu_5}\, p_4^{\mu_1}\, p_4^{\mu_2}\,, \qquad &
T_{32}^{\mu_1,...,\mu_5} = {}& p_2^{\mu_3}\, p_2^{\mu_4}\, p_3^{\mu_5}\, p_4^{\mu_1}\, p_4^{\mu_2}\,.
\end{alignat}
With these tensors we can therefore rewrite~\eqref{eq:ampl5g}  as
\begin{equation}
\mathcal{A}(p_1,...,p_4) = \epsilon_1^{\mu_1}...\epsilon_5^{\mu_5} \, \sum_{j=1}^{32} \mathcal{F}_j\,
T_j^{\mu_1,...,\mu_5}(p_1,...,p_4) + \mathcal{O}(g^{\mu\nu}_{[-2 \epsilon]})\,, \label{eq:ampl5gD4}
\end{equation}
where $\mathcal{O}(g^{\mu\nu}_{[-2 \epsilon]})$ indicates tensor structures which live in the $(-2 \epsilon)$-dimensional
space and do not contribute to helicity amplitudes.

Starting from this tensor, we put to zero all terms proportional to $\mathcal{O}(g^{\mu\nu}_{[-2 \epsilon]})$ and fix the helicities of the five external gluons
in all possible ways by using the spinor-helicity formalism. For every helicity configuration, we then define a rescaled
amplitude $\overline{\mathcal{A}}_{\lambda_1\lambda_2\lambda_3\lambda_4\lambda_5}$ which is
invariant under little group transformations, see eq.~\eqref{eq:littlegroup}.
This can be achieved by dividing the corresponding amplitudes
by a suitable prefactor
$K_{\lambda_1\lambda_2\lambda_3\lambda_4\lambda_5}$ for the $h_\lambda = 2^5 = 32$ different helicity configurations.
For the helicity configurations which are zero at tree-level in QCD we choose
\begin{align}
K_{+++++} &= \frac{s_{12}^2}{\langle 12 \rangle \langle 23 \rangle \langle 34 \rangle \langle 45 \rangle \langle 51 \rangle}\,,
\qquad
K_{-++++} = \frac{\left( \langle 12 \rangle [23] \langle 31 \rangle \right)^2}{\langle 12 \rangle \langle 23 \rangle \langle 34 \rangle \langle 45 \rangle \langle 51 \rangle}
\label{eq:scales1}
\end{align}
and cyclic permutations thereof.
For the MHV amplitudes, instead, we can choose the tree-level
Parke-Taylor amplitudes as rescaling factor, e.g.
\begin{align}
K_{--+++} = \frac{\langle 12 \rangle^4 }{\langle 12 \rangle \langle 23 \rangle \langle 34 \rangle \langle 45 \rangle \langle 51 \rangle}\,,
\label{eq:scales2}
\end{align}
and similarly for the remaining $9$ configurations. Scaling factors
for the helicity configurations with three or more negative helicities
can be obtained by complex conjugation of eqs.~(\ref{eq:scales1},\ref{eq:scales2}).

Before deriving the helicity projectors, it is convenient to obtain a rational parametrisation of the spinor products $\langle ij\rangle$, $[ij]$
and of the external invariants $s_{ij}$, since this avoids the need of
performing tedious spinor algebra.  For the case of five massless external particles, we can use the parametrisation
in terms of momentum twistors~\cite{Hodges:2009hk} provided in~\cite{Badger:2017jhb}.
We define a momentum twistor $Z_j$ for each momentum and write the parametrisation in matrix form as
\begin{equation}
Z = \begin{pmatrix} 1 & 0 & \frac{1}{x_1} & \frac{1+x_2}{x_1 x_2}  &  \frac{1+x_3 (1+x_2)}{x_1 x_2 x_3}  \\
                                0 & 1 & 1 & 1 & 1 \\
                                0 & 0 & 0 & \frac{x_4}{x_2} & 1 \\
                                0 & 0 & 1 & 1 & \frac{x_4 - x_5}{x_4}\end{pmatrix}\,,
\end{equation}
\vspace{0.2cm}
where the $x_j$ are momentum twistor variables. The kinematic invariants can the be written  as
\begin{align}
& s_{12} = x_1\,, \quad s_{23} = x_1 x_4\,, \quad s_{34} = x_1(x_4 + x_3 x_4 - x_2 x_3 + x_2 x_3 x_5)/x_2 \nonumber \\
& s_{45} = x_1 x_5 \,, \qquad s_{51} = x_1 x_3 (x_2 - x_4 + x_5)\,.\label{eq:sijtoxj}
\end{align}
Similarly, for the parity-odd invariant we find
\begin{equation}
{\rm tr}_5 = - x_1^2 \left(x_3 \left(x_5-1\right) x_2^2+\left(2 x_3+1\right) x_4 x_2-\left(x_3+1\right) x_4 \left(x_4-x_5\right)\right)/x_2. \label{eq:tr5toxj}
\end{equation}
An explicit parametrisation of the spinor components in terms of these
variables is given in eq.~(5.10) of~\cite{Peraro:2016wsq} (see also
ref.~\cite{Badger:2016uuq} for a generalisation to other processes).

For each helicity configuration, we now proceed with defining the
functions $R^{\lambda_1,...,\lambda_5}_j$, see eq.~\eqref{eq:Rj}, as
rational functions of the momentum twistor variables.  As pointed out
at the end of section~\ref{sec:idea}, we may now choose to continue
using the variables $x_j$ or alternatively switch back to Mandelstam
invariants.  In this example we pick the latter option.  It is
straightforward to invert the relations in
eqs. (\ref{eq:sijtoxj},\ref{eq:tr5toxj}) and write
\begin{equation}
  R^{\lambda_1,...,\lambda_5}_j = R^{\lambda_1,...,\lambda_5}_{+,j}(s_{ij}) + \textrm{tr}_5\, R^{\lambda_1,...,\lambda_5}_{-, j}(s_{ij}),
\end{equation}
where $R^{\lambda_1,...,\lambda_5}_{\pm,j}$ are rational functions of
the Mandelstam invariants $s_{ij}$.  Notice that this representation
is unique.

Having defined the functions $R^{\lambda_1,...,\lambda_5}_j$, we are
now ready to reconstruct our physical projectors, defined as in
eq.~\eqref{eq:helproj}. We use \textsc{FiniteFlow}~\cite{Peraro:2016wsq} to invert the
$32 \times 32$ matrix and reconstruct directly the physical projectors as linear combinations of the
$32$ tensors in~\eqref{eq:tensors5g}
\begin{equation}
\mathcal{P}_{\lambda_1,...,\lambda_5}
= \epsilon_{1\, \mu_1}^*...\epsilon_{5\, \mu_5}^*\, \sum_{k=1}^{32} \mathcal{C}^{\lambda_1,\ldots,\lambda_5}_{k} T_k^{\mu_1,...,\mu_5}.
\label{eq:helproj25g}
\end{equation}
Similarly to the coefficients $R^{\lambda_1,...,\lambda_5}_j$ above, we can write the
$\mathcal{C}^{\lambda_1,\ldots,\lambda_5}_k$ in
eq.~\eqref{eq:helproj25g} as
\begin{equation}
  \mathcal{C}^{\lambda_1,\ldots,\lambda_5}_{k} = \mathcal{C}^{\lambda_1,\ldots,\lambda_5}_{+,k} + \textrm{tr}_5\, \mathcal{C}^{\lambda_1,\ldots,\lambda_5}_{-,k}
\end{equation}
where the parity even and odd parts
$\mathcal{C}^{\lambda_1,\ldots,\lambda_5}_{+,k}$ and
$\mathcal{C}^{\lambda_1,\ldots,\lambda_5}_{-,k}$ are rational
functions of the Mandelstam invariants $s_{ij}$ and are only
determined by $R^{\lambda_1,...,\lambda_5}_{+,j}$ and
$R^{\lambda_1,...,\lambda_5}_{-,j}$ respectively.  Explicit
expressions for the coefficients
$\mathcal{C}^{\lambda_1,\ldots,\lambda_5}_{\pm,k}$ for a full set of
helicity configurations are given in ancillary files.  As exemplification, we write down explicitly the coefficients
of the parity-even part of the projector on the all-plus helicity amplitude.
By defining
$$\mathcal{C}_{+,k}^{+++++} = \frac{4  s_{23} s_{34} s_{45} s_{51}}{\sqrt{2}\, s_{12} \Delta(p_1,p_2,p_4,p_4)^2}\,
 \overline{\mathcal{C}}_{+,k}^{+++++}$$
 where $\Delta(p_1,p_2,p_4,p_4)$ is the Gram-determinant of the four momenta
  $$\Delta(p_1,p_2,p_4,p_4) = \left(-s_{23} s_{34}+s_{12} \left(s_{23}-s_{51}\right)+s_{45} \left(s_{34}+s_{51}\right)\right){}^2+4 s_{34} \left(s_{12}+s_{23}-s_{45}\right) s_{45}
   s_{51}\,,$$
we find
{\small
\begin{align}
\overline{\mathcal{C}}_{+,1}^{+++++} &=
\left(s_{12}+s_{23}-s_{34}-s_{45}\right)  \left(s_{23}+s_{34}-s_{51}\right){}^2, \nonumber \\
\overline{\mathcal{C}}_{+,2}^{+++++} &=
\left(s_{23}+s_{34}-s_{51}\right) \left(s_{23}^2-\left(s_{45}+s_{51}\right) s_{23}
+s_{12} \left(s_{23}-s_{51}\right)+\left(s_{34}+s_{45}\right) s_{51}\right), \nonumber \\
\overline{\mathcal{C}}_{+,3}^{+++++} &=
\left(s_{23}+s_{34}-s_{51}\right){}^2
   \left(s_{12}-s_{34}+s_{51}\right), \nonumber \\
\overline{\mathcal{C}}_{+,4}^{+++++} &=
\left(s_{23}+s_{34}-s_{51}\right) \left(s_{12} \left(s_{23}-s_{51}\right)
+\left(s_{23}+s_{34}-s_{51}\right) s_{51}\right), \nonumber \\
\overline{\mathcal{C}}_{+,5}^{+++++} &=
\left(s_{12}+s_{23}-s_{45}\right) \left(s_{23}+s_{34}-s_{51}\right)
   \left(s_{23}-s_{45}-s_{51}\right), \nonumber \\
\overline{\mathcal{C}}_{+,6}^{+++++} &=
\left(s_{12}+s_{23}-s_{45}\right) \left(s_{23}+s_{34}-s_{51}\right) \left(s_{23}-s_{45}-s_{51}\right), \nonumber \\
\overline{\mathcal{C}}_{+,7}^{+++++} &=
-\left(s_{23}+s_{34}-s_{51}\right) \left(s_{12}-s_{34}+s_{51}\right)
   \left(-s_{23}+s_{45}+s_{51}\right), \nonumber \\
\overline{\mathcal{C}}_{+,8}^{+++++} &=
\left(s_{23}+s_{34}-s_{51}\right) \left(s_{23}-s_{45}-s_{51}\right) \left(s_{12}+s_{51}\right), \nonumber \\
\overline{\mathcal{C}}_{+,9}^{+++++} &=
\left(s_{12}^2+\left(s_{23}-s_{34}-s_{45}\right) s_{12}+s_{23} \left(s_{23}-s_{45}-s_{51}\right)\right)
   \left(s_{23}+s_{34}-s_{51}\right), \nonumber \\
\overline{\mathcal{C}}_{+,10}^{+++++} &=
\left(s_{23}-s_{51}\right) s_{12}^2+\left(2 s_{23}^2-2 \left(s_{45}+s_{51}\right)
s_{23}+\left(s_{34}+s_{45}\right) s_{51}\right) s_{12}+s_{23} \left(-s_{23}+s_{45}+s_{51}\right){}^2, \nonumber \\
\overline{\mathcal{C}}_{+,11}^{+++++} &=
s_{12}
   \left(s_{23}+s_{34}-s_{51}\right) \left(s_{12}-s_{34}+s_{51}\right), \nonumber \\
\overline{\mathcal{C}}_{+,12}^{+++++} &=
s_{12} \left(s_{23}^2-s_{45} s_{23}+s_{12} \left(s_{23}-s_{51}\right)+\left(s_{34}-s_{51}\right) s_{51}\right), \nonumber \\
\overline{\mathcal{C}}_{+,13}^{+++++} &=\left(s_{12}+s_{23}-s_{45}\right) \left(s_{23}+s_{34}-s_{51}\right)
   \left(s_{23}-s_{45}-s_{51}\right), \nonumber \\
\overline{\mathcal{C}}_{+,14}^{+++++} &=
\left(s_{12}+s_{23}-s_{45}\right) \left(s_{23}-s_{45}-s_{51}\right) \left(s_{12}+s_{23}-s_{45}-s_{51}\right), \nonumber \\
\overline{\mathcal{C}}_{+,15}^{+++++} &=
0, \nonumber \\
\overline{\mathcal{C}}_{+,16}^{+++++} &=
s_{12} \left(s_{12}+s_{23}-s_{45}\right) \left(s_{23}-s_{45}-s_{51}\right), \nonumber \\
\overline{\mathcal{C}}_{+,17}^{+++++} &=
s_{23}
   \left(s_{12}+s_{23}-s_{45}\right) \left(s_{23}+s_{34}-s_{51}\right), \nonumber \\
\overline{\mathcal{C}}_{+,18}^{+++++} &=
s_{23} \left(s_{12}+s_{23}-s_{45}\right) \left(s_{23}+s_{34}-s_{51}\right), \nonumber \\
\overline{\mathcal{C}}_{+,19}^{+++++} &=
s_{23} \left(s_{23}+s_{34}-s_{51}\right) \left(s_{12}-s_{34}+s_{51}\right), \nonumber \\
\overline{\mathcal{C}}_{+,20}^{+++++} &=
s_{23} \left(s_{23}+s_{34}-s_{51}\right) \left(s_{12}+s_{51}\right), \nonumber \\
\overline{\mathcal{C}}_{+,21}^{+++++} &=
-\left(s_{12}+s_{23}-s_{45}\right) \left(-s_{23}^2+\left(-s_{34}+s_{45}+s_{51}\right)
s_{23}+s_{34} \left(s_{12}-s_{34}+s_{51}\right)\right), \nonumber \\
\overline{\mathcal{C}}_{+,22}^{+++++} &=
\left(s_{23}+s_{34}\right)
   \left(s_{12}+s_{23}-s_{45}\right) \left(s_{23}-s_{45}-s_{51}\right), \nonumber \\
\overline{\mathcal{C}}_{+,23}^{+++++} &=
-\left(s_{12}-s_{34}+s_{51}\right) \left(-s_{23}^2+\left(s_{45}+s_{51}\right) s_{23}+s_{12} s_{34}\right), \nonumber \\
\overline{\mathcal{C}}_{+,24}^{+++++} &=
\left(s_{23}-s_{45}-s_{51}\right) \left(s_{12} \left(s_{23}+s_{34}\right)+s_{23}
   s_{51}\right), \nonumber \\
\overline{\mathcal{C}}_{+,25}^{+++++} &=
s_{23} \left(s_{12}+s_{23}-s_{45}\right) \left(s_{23}+s_{34}-s_{51}\right), \nonumber \\
\overline{\mathcal{C}}_{+,26}^{+++++} &=
s_{23} \left(s_{12}+s_{23}-s_{45}\right) \left(s_{12}+s_{23}-s_{45}-s_{51}\right), \nonumber \\
\overline{\mathcal{C}}_{+,27}^{+++++} &=
0, \nonumber \\
\overline{\mathcal{C}}_{+,28}^{+++++} &=
s_{12} s_{23} \left(s_{12}+s_{23}-s_{45}\right), \nonumber \\
\overline{\mathcal{C}}_{+,29}^{+++++} &=
-\left(s_{12}+s_{23}-s_{45}\right){}^2 \left(s_{12}-s_{23}-s_{34}+s_{51}\right), \nonumber \\
\overline{\mathcal{C}}_{+,30}^{+++++} &=
\left(s_{12}+s_{23}-s_{45}\right){}^2 \left(s_{23}-s_{45}-s_{51}\right), \nonumber \\
\overline{\mathcal{C}}_{+,31}^{+++++} &=
-s_{12} \left(s_{12}+s_{23}-s_{45}\right)
   \left(s_{12}-s_{34}+s_{51}\right), \nonumber \\
\overline{\mathcal{C}}_{+,32}^{+++++} &=s_{12} \left(s_{12}+s_{23}-s_{45}\right) \left(s_{23}-s_{45}-s_{51}\right)\,.
\end{align}
}

As a check of the consistency of our approach, we can obtain the same
result starting from a full $d$-dimensional tensor decomposition.
In particular, we could ignore the fact that $g^{\mu \nu}$ is not
linearly independent and decide to start from the full $d$-dimensional
tensor in eq.~\eqref{eq:ampl5g}.
If we do so, we can  formally write the physical helicity projectors as linear
combinations of the original 142 $d$-dimensional projectors.   We then use \textsc{FiniteFlow}~\cite{Peraro:2016wsq}
to invert the corresponding $142\times 142$ matrix in eq.~\eqref{eq:systeminverse} numerically and
use this to reconstruct only the physical projectors directly in terms of the
original tensor structures $T_j^{\mu_1,...,\mu_5}$, as we did in
eq.~\eqref{eq:helproj2}.  The analytic reconstruction takes a couple of minutes on a
modern laptop. As a result, as expected, we find that all the coefficients which multiply the
tensors which depend on $g^{\mu \nu}$ turn out to be zero and we can recover the very same
result discussed above. Clearly, by removing the $(-2\epsilon)$-dimensional tensors from the beginning, all manipulations
are much simpler and reconstruction procedure runs through only in
a few seconds.

\subsection{Five-gluon scattering at one-loop in QCD}
As a validation of the helicity projectors newly derived, we have used them to compute
 the one-loop corrections to five-gluon scattering in QCD. While
this calculation is rather simple using modern one-loop techniques based on either generalised
unitarity or integrand reduction, see for example ref.~\cite{Bern:1993mq},  it would clearly constitute
a challenge using standard projection-based techniques.

Following common practice in these calculations, we decompose the tree-level and one-loop
five-gluon helicity amplitudes
in terms of coloured-ordered primitive amplitudes.
\comment{Also, as described above, we always
considered little-group invariant helicity amplitudes. We start writing
\begin{equation}
\overline{\mathcal{A}}_{\lambda_1...\lambda_5}(p_1,...,p_5) = \overline{\mathcal{A}}^{(0)}_{\lambda_1...\lambda_5}(p_1,...,p_5) +
\left(\frac{\alpha_s}{2 \pi} \right)\overline{\mathcal{A}}^{(1)}_{\lambda_1...\lambda_5}(p_1,...,p_5) + \mathcal{O}(\alpha_s^2)
\end{equation}
and decompose the tree-level and one-loop helicity amplitudes in primitive amplitudes as follows
\begin{align}
\overline{\mathcal{A}}^{(0)}_{\lambda_1...\lambda_5}(p_1,...,p_5) &= \sum_{\sigma=1}^{12} \mathcal{A}^{(0),\sigma}_{\lambda_1...\lambda_5}\, \mathcal{T}^{[1]}_\sigma \nonumber \\
\overline{\mathcal{A}}^{(1)}_{\lambda_1...\lambda_5}(p_1,...,p_5) &= \sum_{\sigma=1}^{12} \mathcal{A}^{(1),\sigma}_{\lambda_1...\lambda_5}\, \mathcal{T}^{[1]}_\sigma
+ \sum_{\sigma=13}^{22} \mathcal{A}^{(1),\sigma}_{\lambda_1...\lambda_5}\, \mathcal{T}^{[2]}_\sigma \label{eq:primampl}
\end{align}
where
\begin{align}
\mathcal{T}^{[1]}_1 = {\rm Tr}(T^{a_1}T^{a_2} T^{a_3} T^{a_4} T^{a_5})\,,\quad \mathcal{T}^{[2]}_{13} = {\rm Tr}(T^{a_1}T^{a_2} ) {\rm Tr}(T^{a_3} T^{a_4} T^{a_5})\,,
\end{align}
and the remaining ones are given by all permutations of the indices $a_1,...,a_5$, modulo cyclic permutations within the individual traces.
The rescaled helicity amplitudes are defined dividing by the factors in eqs.(\ref{eq:scales1},\ref{eq:scales2}).
}
It is well known that all one-loop primitive amplitudes
can be obtained from the coefficient of the colour structure
${\rm Tr}(T^{a_1}T^{a_2} T^{a_3} T^{a_4} T^{a_5})$ of the following
helicity configurations
\begin{equation*}
\mathcal{A}_{+++++}\,, \quad \mathcal{A}_{-++++}\,, \quad \mathcal{A}_{--+++}\,,\quad  \mathcal{A}_{-+-++}\, .
\end{equation*}
\comment{
where we used the fact that the only non-zero amplitudes at tree-level are the MHV amplitudes and
 that, only up to one loop, the double-trace primitive amplitudes are not independent.
All other primitive helicity amplitudes can be obtained by complex conjugation and cyclic permutations of the external gluons.
Finally, in QCD at the one-loop order the individual primitive amplitudes in eqs.~\eqref{eq:primampl} can be further decomposed as
\begin{equation}
 \mathcal{A}^{(1),\sigma}_{\lambda_1...\lambda_5} = N_c \, \mathcal{B}^{(1),\sigma}_{\lambda_1...\lambda_5} + N_f \, \mathcal{C}^{(1),\sigma}_{\lambda_1...\lambda_5}\,,
\end{equation}
where $N_c$ is the number of colours and $N_f$ the number of fermions. We ignore this further decomposition here since it is straightforward
to process the entire amplitude at once.
} 

In order to compute these amplitudes, we first generate all relevant Feynman diagrams with QGRAF~\cite{Nogueira:1991ex}
and sort them selecting only the ones corresponding to the relevant colour-ordered amplitudes.
We then proceed by applying on each diagrams the projectors defined in eq.~\eqref{eq:helproj25g}. In practice, we
prefer to compute for each diagram the $32$  contractions with the $32$ tensors in eq.~\eqref{eq:tensors5g} independently.
More explicitly, for every Feynman diagram $\mathcal{D}_j$, we extract the gluon polarisation vectors
$$\mathcal{D}_j = (\epsilon_{1\, \mu_1}...\epsilon_{5\, \mu_5}) D_j^{\mu_1,...,\mu_5}\,,$$
and compute the quantity
\begin{equation}
\mathcal{D}_{jk} = \sum_{pol} \left( \epsilon_{1\, \mu_1}^*...\epsilon_{5\, \mu_5}^* \epsilon_{1\, \nu_1}...\epsilon_{5\, \nu_5}  \right) T_k^{\mu_1,...,\mu_5} \,   D_j^{\nu_1,...,\nu_5}\,.
\end{equation}
Due to the transversality and gauge constraints that we imposed on the
tensor structures, our polarisations sums are given by
\begin{align}
\sum_{pol}\, \epsilon_{1,\mu_1}^* \, \epsilon_{1\, \nu_1} =
- g_{\mu_1 \nu_1} + \frac{p_{1,\mu_1} p_{2,\nu_1} + p_{2,\mu_1} p_{1,\nu_1}}{p_1 \cdot p_2}\,,
\end{align}
and cyclic permutations thereof.
Once all $\mathcal{D}_{jk}$ have been computed, the relevant helicity amplitudes can be computed by summing all Feynman diagrams and
assembling them together as in eq.~\eqref{eq:helproj25g}. While all these manipulations could be performed
efficiently using \textsc{FiniteFlow}~\cite{Peraro:2019svx},
the simplicity of the tensor structures and of the helicity projectors allow us to perform them using FORM~\cite{Vermaseren:2000nd}
and Reduze~\cite{vonManteuffel:2012np} in few hours
on a laptop.  In our calculation we have included the full dependence on the number
of colours $N_c$ and the number of fermions $N_f$  and we have verified explicitly that our unrenormalised
helicity amplitudes agree with known results, even before substituting
the explicit analytical results for the master integrals~\cite{privcom}.



\section{Conclusions}
\label{sec:conclusions} \setcounter{equation}{0}
\numberwithin{equation}{section}

We presented an efficient method for building physical projector operators for
helicity amplitudes, which is suitable for applications to multi-leg processes.  While it is common belief that a
projector-based approach to compute multi-loop multi-leg scattering amplitudes in perturbative QFT
becomes soon impractical due to the proliferation of the number of tensor structures and of the complexity of the
corresponding projectors required, in this paper we have shown that this is not necessarily the case.
In particular, we have demonstrated that  if one aims to build projection operators that reconstruct only
physical helicity amplitudes, huge simplifications take place due the large redundancy of the generic
$d$-dimensional tensor structure.
It turns out that, when considering the scattering of $n\geq5$ particles of arbitrary spin, the number of different
helicity amplitudes $h_\lambda$ provides a higher bound for the number of different tensor structures
that are required in order to reconstruct them.  Hence,  in these cases, one can obtain
a full set of independent helicity amplitudes from the contraction of
the amplitude with no more than $h_\lambda$ tensor structures.   Moreover, the corresponding
projection operators turn out to be substantially simpler.

Starting from $n=5$ external legs, this method yields additional
drastic simplifications compared to traditional projector-based
approaches. Indeed, in this case the entire four-dimensional space can
be spanned by the four independent external momenta and all tensor
structures which involve the metric tensor $g^{\mu \nu}$ turn out to
be redundant.  We have demonstrated this explicitly by studying the
tensor decomposition for five-gluon scattering in QCD and comparing
the standard $d$-dimensional approach with our method.  We have found
that, while in the standard approach there are $142$ independent
tensor structures and thus $142$ rather cumbersome projection
operators, their number drops to $32$ when projecting directly on the
independent helicity amplitudes. As expected, $32$ is also the number
of different helicity configurations $2^5 = 32$.  We derived
explicitly the helicity projection operators, which are attached to
the arXiv submission of this paper, and we validated them by computing
the tree-level and one-loop corrections to five-gluon scattering in
QCD.  The simplifications obtained in this way were so substantial
that the whole calculation could be completed in few hours on a laptop
computer with a straightforward application of standard computer
algebra systems.  Similar simplifications were observe when computing
projector operators for other five-point processes, such as the
scattering of four gluons and a (massive) scalar.

While recent progress
in integrand reduction and generalised unitarity has considerably
improved the possibilities of computing multi-leg helicity amplitudes,
enhancing the spectrum of techniques which can tackle these processes
is definitely useful for further progress.  Given the generality,
the relative easy-of-use and the familiarity of projector-based
approaches compared to alternative techniques, we believe that making
them applicable to more complex processes will prove beneficial to
future calculations. We also stress
that, despite being commonly seen as an alternative, integrand reduction may in principle be applied in conjunction
with projection operators.

While this constitutes a very interesting development in view of the calculations of two-loop corrections to
other processes which involve five particles in the final state, we note that one should expect
even more substantial simplifications when considering even more particles in the final state.
Clearly, we are well aware of the fact that these calculations are extremely complicated irrespective of the approach used.
Nevertheless, we believe that this
paper provides an important contribution to substantially simplify one of the computationally most demanding
steps required.


\section*{Acknowledgements}
We thank Simon Badger, Claude Duhr and Thomas Gehrmann for inspiring discussions about the
ideas presented in this paper.  We are grateful to Simon Badger, Christian Broennum-Hansen and
Heribertus Bayu Hartanto for providing analytic expressions of one-loop five-gluon helicity amplitudes in terms of
master integrals, which we used
to check our calculation.
This
project has received funding from the European Union's Horizon 2020
research and innovation programme under the Marie Sklodowska-Curie
grant agreement 746223 and from the ERC grant 637019 ``MathAm''.

\appendix

\bibliographystyle{bibliostyle}
\bibliography{Biblio}
\end{document}